\documentclass[prc,preprint,showpacs,tightenlines,%
                floatfix,amsfonts,nofootinbib]{revtex4}

\usepackage{graphicx}
\usepackage{bm}
\usepackage{amssymb}

\begin{document}

%
%
%
%
\newcommand{\covdev}{{\widetilde \partial}}
\newcommand{\finalnewpage}{\newpage}
\newcommand{\GEp}{G_{Ep}}
\newcommand{\GEn}{G_{En}}
\newcommand{\GMp}{G_{Mp}}
\newcommand{\GMn}{G_{Mn}}
\newcommand{\lagrang}{{\cal L}}
\newcommand{\newg}{{\skew2\overline g}_\rho}
\newcommand{\Nbar}{\skew3\overline N \mkern2mu}
\newcommand{\stroke}[1]{\mbox{{$#1$}{$\!\!\!\slash \,$}}}
\newcommand{\Ximinus}{\xi \frac{\tau_3}{2} \xi^\dagger%
                     - \xi^\dagger \frac{\tau_3}{2} \xi}
\newcommand{\Xiplus}{\xi \frac{\tau_3}{2} \xi^\dagger%
                     + \xi^\dagger \frac{\tau_3}{2} \xi}
\newcommand{\Xiaminus}{\xi \frac{\tau_a}{2} \xi^\dagger%
                     - \xi^\dagger \frac{\tau_a}{2} \xi}
\newcommand{\Xiaplus}{\xi \frac{\tau_a}{2} \xi^\dagger%
                     + \xi^\dagger \frac{\tau_a}{2} \xi}

%
%
\def\dthree#1{\intback{\rm d}^3\intback #1}
\def\dthreex{\dthree{x}}
\def\fpi{f_{\pi}}
\def\gammafive{\gamma^5}
\def\gammafivel{\gamma_5}
\def\gammamu{\gamma^{\mu}}
\def\gammamul{\gamma_{\mu}}
\def\intback{\kern-.1em}
\def\kfermi{k_{\sssize {\rm F}}}    
\def\mn{{\mu\nu}}
\def\psibar{\overline\psi}
\def\sigmamunu{\sigma^\mn}
\def\sigmamunul{\sigma_\mn}
\def\Tr{\mathop{\rm Tr}\nolimits}

\let\dsize=\displaystyle
\let\tsize=\textstyle
\let\ssize=\scriptstyle
\let\sssize\scriptscriptstyle
%
%
%


\newcommand{\beq}{\begin{equation}}
\newcommand{\eeq}{\end{equation}}
\newcommand{\beqa}{\begin{eqnarray}}
\newcommand{\eeqa}{\end{eqnarray}}

\def\Inthelimit#1{\lower1.9ex\vbox{\hbox{$\
   \buildrel{\hbox{\Large \rightarrowfill}}\over{\scriptstyle{#1}}\ $}}}

\title{A Field-Theoretic Parametrization of \\
       Low-Energy Nucleon Form Factors}

\author{Brian D. Serot}\email{serot@indiana.edu}
\affiliation{Department of Physics and Nuclear Theory Center
             Indiana University, Bloomington, IN\ \ 47405}

%
\author{\null}
\noaffiliation

%
\date{\today\\[20pt]}

\begin{abstract}
A field-theoretic parametrization is proposed for nucleon
electromagnetic form factors at momentum transfer less than 600 MeV.
The parametrization is part of a larger effective field theory
lagrangian that is Lorentz covariant and chiral symmetric, and that
has been used to successfully describe bulk and single-particle
properties of medium to heavy mass nuclei.  The parametrization is
based on vector meson dominance and a derivative expansion of
nucleon couplings to the electromagnetic fields.  At lowest order in
the expansion, it is possible to fit all four parameters to modern
data on the rms radii of the nucleon form factors. At
next-to-leading order it is possible to fit the form factors to
within a few percent up to momentum transfers of 600 MeV.  The
vector meson dominance contributions are crucial in this fit, since
a simple expansion in powers of momentum transfer would require
many, many terms to achieve comparable accuracy. The ability to fit
single-nucleon form factors up to 600 MeV momentum transfer makes
possible the study of two-body electromagnetic exchange currents
within this effective field theory framework.

\end{abstract}

\smallskip

\maketitle

\section{Introduction}
\label{sec:intro}

Lorentz-covariant meson--baryon effective field theories of the
nuclear many-body problem (often called \emph{quantum hadrodynamics}
or QHD) have been known for many years to provide a realistic
description of the bulk properties of nuclear matter and heavy
nuclei. (For reviews, see
Refs.~\cite{SW86,Reinhard89,Gambhir90,BDS92,Ring96,SW97}.) Recently,
a QHD effective field theory (EFT) has been proposed
\cite{FST97,FSp00,FSL00,EvRev00,LNP641,EMQHD07} that includes all
the relevant symmetries of the underlying QCD. In particular, the
spontaneously broken $SU(2)_L \times SU(2)_R$ chiral symmetry is
realized nonlinearly. The motivation for this EFT and some
calculated results are discussed in Refs.~\cite{SW97,FST97,HUERTAS02,HUERTASwk,%
HUERTAS04,MCINTIRE04,MCINTIRE05,JDW04,MCINTIRE07,HU07,MCINTIRE08}.

This QHD EFT has three desirable features: (1) It uses the same
degrees of freedom to describe the currents and the
strong-interaction dynamics; (2) It respects the same internal
symmetries, both discrete and continuous, as the underlying QCD
(before and after electromagnetic interactions are included); and
(3) Its parameters can be calibrated using strong-interaction
phenomena, like $\pi$N scattering and the empirical properties of
finite nuclei (as opposed to electroweak interactions with nuclei).
It thus provides a natural framework, based on a single lagrangian,
for discussing the roles of one-body and two-body currents in
nuclear electromagnetic interactions.

The nucleon electromagnetic (EM) structure (form factors) is
described in this EFT using a combination of vector meson dominance
(VMD) \cite{FST97,SAKURAI60,GMZ,GMSW62,KLZ,BERNSTEIN68,SAKURAI69}
and a derivative expansion for nucleon interactions with the EM
field. In the applications of this EFT to nuclear structure noted
above, however, only the lowest-order derivative couplings were
included, so that the form factors provided an accurate description
of the single-nucleon electron scattering data only up to roughly
250 MeV momentum transfer.  In contrast, if one is to study
\emph{two-body} (exchange) currents, one must reproduce the
single-nucleon form factors accurately up to at least 600 MeV
momentum transfer, where two-body contributions are expected to be
visible.  This momentum scale should be accessible in this
low-energy hadronic EFT \cite{FST97,RF97,FSp00}.

Our motivation for this study of nucleon form factors is twofold.
First, we want to update the lowest-order fits of Ref.~\cite{FST97}
to include the large amount of low-energy, high-precision data that
became available in the early 2000's.  Second, we extend the fits to
the next order in momentum transfer and show that the form factors
will accurately reproduce the empirical results up to roughly 600
MeV momentum transfer.  This will make them suitable for studies of
exchange currents within the QHD EFT.

In the past ten or fifteen years, much new data on the nucleon EM
form factors have been obtained using both unpolarized electron
scattering and polarization transfer. (For recent reviews, see
Refs.~\cite{GAO03,ARRINGTON07}.)  There have also been numerous
attempts at fitting the improved data set; for example, see
Refs.~\cite{KELLY02,KELLY04,ALBERICO09}.  For the present study, we
are most interested in the work of Kelly \cite{KELLY04}, who
achieved excellent fits with a small number of free parameters.  In
particular, the fits are good enough over the momentum transfer
range of interest to us that we will simply fit our EFT parameters
to Kelly's analytic results rather than to the data itself.  Since
our best fits reproduce Kelly's at the few percent level, this
procedure is justified.

One of our interesting results is that a straightforward $Q^2$
expansion of Kelly's analytic results is inadequate unless many,
many terms are retained.  (Here $Q^2 \equiv -q^2$ is the square of
the spacelike four-momentum transfer.)  The presence of the VMD
contributions in the EFT approach greatly improves the situation.
Moreover, it is important to include the new EFT parameters in such
a way that the error is minimized for the whole relevant range of
$Q^2$, not just $Q^2 \to 0$.

This paper is organized as follows: In Sec.~II, we return to the
lowest-order parametrization of Ref.~\cite{FST97} and re-fit the EFT
parameters to the new data set.  This allows us to determine
mean-square radii for all four form factors
(neutron/proton--electric/magnetic). In Sec.~III, we extend the EFT
lagrangian by introducing new parameters and use them to fit the
higher-momentum transfer behavior of the form factors.  Sec.~IV is a
brief Summary.

\section{Re-fit of Lowest-Order Parameters}
\label{sec:refit}

In this section, we consider the form factors as described in
Ref.~\cite{FST97}.  We follow the conventions of
Refs.~\cite{FST97,EMQHD07}. Rather than work with the Dirac ($F_1$)
and Pauli ($F_2$) form factors, defined in terms of the nucleon EM
vertex as
\beq
\Gamma^\mu = F_1 (Q^2) \gammamu + F_2 (Q^2) \frac{i \sigmamunu
q_\nu}{2 M} \ ,
\label{eq:DiracFF}
\eeq
where $M$ is the nucleon mass and $F_2$ contains the anomalous
magnetic moment, here we will primarily concentrate on the Sachs
form factors
\beq
G_E (Q^2) = F_1 (Q^2) - \tau F_2 (Q^2)\ ,
\quad G_M (Q^2) = F_1
(Q^2) + F_2 (Q^2) \ , \label{eq:Sachsdef} \eeq
where $\tau \equiv Q^2/4M^2 \equiv -q^2/4M^2$ in terms of the
four-momentum $q^\mu$, and we have not distinguished the charge
states. The charge states are written in terms of the isoscalar
$(0)$ and isovector $(1)$ parts as, for example,
\beq
F_p = F^{(0)} + F^{(1)}\ , \quad
F_n = F^{(0)} - F^{(1)} \ .
\label{eq:chargedef}
\eeq

The simple parametrizations used by Kelly \cite{KELLY04} take the
form
\beq G (Q^2) = \frac{\sum_{k \,=\, 0}^{n} a_k \tau^k}{1 + \sum_{k
\,=\, 1}^{n+2} b_k \tau^k} \ ,
\label{eq:KELLYdef}
\eeq
which guarantees the correct asymptotic dependence at large $Q^2$:
$G (Q^2) \propto Q^{-4}$.  This will not concern us, as we are
interested in parametrizing the form factors at small $Q^2$.  With
$n=1$ and $a_0 =1$, this parametrization gives excellent fits to
$\GEp$, $\GMp / \mu_p$, and $\GMn / \mu_n$ (where $\mu_i$ is the
full magnetic moment) using four parameters each \cite{KELLY04}. For
$\GEn$, Kelly follows the so-called Galster parametrization
\cite{GALSTER71}:
\beq
\GEn (Q^2) = \frac{A \tau}{1 + B \tau} \, G_D (Q^2) \ ,
\label{eq:Galster}
\eeq
where the dipole form factor is
\beq
G_D (Q^2) \equiv \frac{1}{(1 + Q^2/\Lambda^2 )^2} \ , \quad
\Lambda^2 = 0.71\,\mathrm{GeV}^2 \ ,
\label{eq:Dipole}
\eeq
and $A$ and $B$ are fitted parameters.

For our parametrization, we use set Q2 of Ref.~\cite{FST97}.  This
provides an accurate fit to bulk and single-particle nuclear
properties and leaves the anomalous coupling to the isoscalar vector
meson (``omega'') undetermined; we will determine it here.  We will
need the mass parameters
\beqa M &=& 939\,\mathrm{MeV} = 4.7585\,\mathrm{fm}^{-1} \ ,
\nonumber \\[3pt] m_v &=& 782\,\mathrm{MeV} = 3.963\,\mathrm{fm}^{-1} \ ,
\nonumber \\[3pt] m_{\rho} &=& 770\,\mathrm{MeV} =
3.902\,\mathrm{fm}^{-1} \ ,
\label{eq:massparams}
\eeqa
the anomalous magnetic moments
\beq \lambda_p = 1.793 \ , \quad \lambda_n = -1.913\ ,
\label{eq:anomalous}
\eeq
the couplings in Table I, and the electromagnetic coupling $g_\gamma
= 5.0133$, which follows from the decay width $\Gamma_{\rho^0
\,\to\, e^+ e^-} = 6.8\,\mathrm{keV}$.

\begin{table}[t]
\caption{Coupling parameters from set Q2 \protect\cite{FST97}. Note
that the nucleon couplings to the omega and rho mesons ($g_v$ and
$g_\rho$) are determined from the empirical properties of nuclei.
\protect\label{tab:Q2params}} \vspace{0.15in}
     \begin{tabular}{cccccccccccc}
     \hline\hline
\hspace{0.3in}& $\beta^{(0)}$ &\hspace{0.3in}&
\hspace{0.3in}$\beta^{(1)}$\hspace{0.3in}
&\hspace{0.3in}&\hspace{0.3in} $g_v$\hspace{0.3in} &\hspace{0.3in}&
\hspace{0.3in}$g_\rho$\hspace{0.3in} &\hspace{0.3in}& $f_v$
\hspace{0.3in}&\hspace{0.3in}\hspace{0.3in}&
$f_\rho$\hspace{0.3in}\\
\hline \vspace{2pt} &$0.01181$ && $-0.1847$ && $12.2148$ && $8.5572$
&&0&&
$4.264$\\[2pt]
\hline
     \end{tabular}
%
%
\end{table}

We now proceed with the fits to the new data.  The calculations use
the lagrangian of Ref.~\cite{FST97} and are performed at tree level.
Starting with the proton electric form factor,
\beqa \GEp (Q^2) &=& F^{(0)}_1 + F^{(1)}_1 - \frac{Q^2}{4M^2}\,
(F^{(0)}_2
+ F^{(1)}_2 ) \nonumber \\[4pt]
&=& 1 - (\beta^{(0)} + \beta^{(1)})
\frac{Q^2}{2M^2} - \frac{g_v}{3 g_\gamma} \left( 1 - \frac{f_v
Q^2}{4M^2} \right) \frac{Q^2}{Q^2 + m_v^2} \nonumber\\[4pt]
&\null& \quad {} -\frac{g_\rho}{2 g_\gamma} \left( 1 - \frac{f_\rho
Q^2}{4M^2} \right) \frac{Q^2}{Q^2 + m_\rho^2} - \lambda_p
\frac{Q^2}{4M^2} \ . \label{eq:GEpold} \eeqa
If we define the mean-square radius as
\beq r^2_{Ep} \equiv -6 \left.\frac{\mathrm{d}\GEp (Q^2)}{\mathrm{d}
Q^2}\right)_{Q^2 \,=\, 0} \ , \label{eq:msrad} \eeq
then
\beqa  r^2_{Ep} &=& \frac{1}{2} \, \left[6 \left(
\frac{\beta^{(0)}}{M^2} + \frac{2 g_v}{3 g_\gamma m^2_v} \right) + 6
\left(\frac{\beta^{(1)}}{M^2} + \frac{g_\rho}{g_\gamma
m^2_\rho}\right) \right] + \frac{3 \lambda_p}{2M^2} \nonumber\\[4pt]
&\equiv& \frac{1}{2} \left( \langle r^2 \rangle_{(0)1} + \langle r^2
\rangle_{(1)1} \right) + \frac{3 \lambda_p}{2M^2} \ ,
\label{eq:rEpsquared} \eeqa
where the mean-square radii on the right-hand side are the isoscalar
and isovector values for the Dirac form factor $F_1$. Inserting the
Q2 parameters leads to
\beq (r^2_{Ep})^{1/2} = 0.862\,\mathrm{fm} \ , \label{eq:rrmsEp}
\eeq
which also agrees with the result in Ref.~\cite{KELLY04}.

Turning now to $\GEn$, we have
\beqa \GEn (Q^2) &=& F^{(0)}_1 - F^{(1)}_1 - \frac{Q^2}{4M^2}\,
(F^{(0)}_2
- F^{(1)}_2 ) \nonumber \\[4pt]
&=&  -(\beta^{(0)} - \beta^{(1)}) \frac{Q^2}{2M^2} - \frac{g_v}{3
g_\gamma} \left( 1 - \frac{f_v
Q^2}{4M^2} \right) \frac{Q^2}{Q^2 + m_v^2} \nonumber\\[4pt]
&\null& \quad {} +\frac{g_\rho}{2 g_\gamma} \left( 1 - \frac{f_\rho
Q^2}{4M^2} \right) \frac{Q^2}{Q^2 + m_\rho^2} - \lambda_n
\frac{Q^2}{4M^2} \ , \label{eq:GEnold} \eeqa
so that
\beqa  r^2_{En} &=& \frac{1}{2} \, \left[6 \left(
\frac{\beta^{(0)}}{M^2} + \frac{2 g_v}{3 g_\gamma m^2_v} \right) - 6
\left(\frac{\beta^{(1)}}{M^2} + \frac{g_\rho}{g_\gamma
m^2_\rho}\right) \right] + \frac{3 \lambda_n}{2M^2} \nonumber\\[4pt]
&=& \frac{1}{2} \left( \langle r^2 \rangle_{(0)1} - \langle r^2
\rangle_{(1)1} \right) + \frac{3 \lambda_n}{2M^2} \ .
\label{eq:rEnsquared} \eeqa
If we set $\langle r^2 \rangle_{(0)1} = \langle r^2 \rangle_{(1)1}$,
as in Ref.~\cite{FST97}, we then find $r^2_{En} =
-0.127\,\mathrm{fm}^2$, in significant disagreement with Kelly's
value of $-0.112 \pm 0.003\,\mathrm{fm}^2$.  We conclude that the
new data shows that
\beq \langle r^2 \rangle_{(0)1} - \langle r^2 \rangle_{(1)1} =
0.0294\,\mathrm{fm}^2 \ .
\label{eq:rsquareddiffone}
\eeq
With this information, together with Eq.~(\ref{eq:rEpsquared}), we
can determine two distinct radii for the Dirac form factor:
\beq \langle r^2 \rangle_{(0)1}^{1/2} = 0.799\,\mathrm{fm} \ , \quad
\langle r^2 \rangle_{(1)1}^{1/2} = 0.780\,\mathrm{fm} \ ,
\label{eq:rmsoneradii}
\eeq
in contrast to the assumption made in Ref.~\cite{FST97}.

We now consider the magnetic form factors, beginning with
\beqa \GMp (Q^2) &=& F^{(0)}_1 + F^{(1)}_1 + F^{(0)}_2 + F^{(1)}_2
\nonumber \\[4pt]
&=& 1 - (\beta^{(0)} + \beta^{(1)} ) \frac{Q^2}{2M^2} - \frac{g_v (1
+ f_v )}{3 g_\gamma}\,\frac{Q^2}{Q^2 + m^2_v} \nonumber\\[4pt]
&\null& \quad {} - \frac{g_\rho (1 + f_\rho )}{2 g_\gamma}\,
\frac{Q^2}{Q^2 + m^2_\rho} + \lambda_p \ . \label{eq:GMpold} \eeqa
As expected, for $Q^2 \to 0$, $\GMp (Q^2) \to 1 + \lambda_p =
\mu_p$.  Thus we normalize $\GMp$ by dividing by $\mu_p$, leading to
the mean-square radius
\beqa  r^2_{Mp} &=& \frac{1}{2 ( 1 + \lambda_p)} \, \left[6 \left(
\frac{\beta^{(0)}}{M^2} + \frac{2 g_v}{3 g_\gamma m^2_v} \right) + 6
\left(\frac{\beta^{(1)}}{M^2} + \frac{g_\rho}{g_\gamma
m^2_\rho}\right) \right] \nonumber\\[4pt]
&\null& \quad {} + \frac{1}{2 ( 1 + \lambda_p)}\left( \frac{4 f_v
g_v}{g_\gamma m^2_v} + \frac{6f_\rho g_\rho}{g_\gamma
m^2_\rho}\right) \nonumber\\[4pt]
&=& \frac{1}{2 ( 1 + \lambda_p)}\left( \langle r^2 \rangle_{(0)1} +
\langle r^2 \rangle_{(1)1} + (\lambda_p + \lambda_n ) \langle r^2
\rangle_{(0)2} + (\lambda_p - \lambda_n ) \langle r^2 \rangle_{(1)2}
\right) \nonumber\\[4pt] &=& 0.7191\,\mathrm{fm}^2 \ , \label{eq:rMpsquared}
\eeqa
where the numerical value is taken from Kelly. Note that the
magnetic radii depend on both the Dirac and Pauli mean-square radii.

For the neutron,
\beqa \GMn (Q^2) &=& F^{(0)}_1 - F^{(1)}_1 + F^{(0)}_2 - F^{(1)}_2
\nonumber \\[4pt]
&=& - (\beta^{(0)} - \beta^{(1)} ) \frac{Q^2}{2M^2} - \frac{g_v (1
+ f_v )}{3 g_\gamma}\,\frac{Q^2}{Q^2 + m^2_v} \nonumber\\[4pt]
&\null& \quad {} + \frac{g_\rho (1 + f_\rho )}{2 g_\gamma}\,
\frac{Q^2}{Q^2 + m^2_\rho} + \lambda_n \ . \label{eq:GMnold} \eeqa
The mean square radius is
\beqa  r^2_{Mn} &=& \frac{1}{2 \lambda_n} \, \left[6 \left(
\frac{\beta^{(0)}}{M^2} + \frac{2 g_v}{3 g_\gamma m^2_v} \right) - 6
\left(\frac{\beta^{(1)}}{M^2} + \frac{g_\rho}{g_\gamma
m^2_\rho}\right) \right] \nonumber\\[4pt]
&\null& \quad {} + \frac{1}{2 \lambda_n}\left( \frac{4 f_v
g_v}{g_\gamma m^2_v} - \frac{6f_\rho g_\rho}{g_\gamma
m^2_\rho}\right) \nonumber\\[4pt]
&=& \frac{1}{2 \lambda_n}\left( \langle r^2 \rangle_{(0)1} - \langle
r^2 \rangle_{(1)1} + (\lambda_p + \lambda_n ) \langle r^2
\rangle_{(0)2} - (\lambda_p - \lambda_n ) \langle r^2 \rangle_{(1)2}
\right) \nonumber\\[4pt] &=& 0.8226\,\mathrm{fm}^2 \ . \label{eq:rMnsquared}
\eeqa

With the results in Eqs.~(\ref{eq:rmsoneradii}),
(\ref{eq:rMpsquared}), and (\ref{eq:rMnsquared}), it is now simple
algebra to determine the rms radii for the Pauli form factor, with
the results
\beq \langle r^2 \rangle_{(0)2}^{1/2} = 1.30\,\mathrm{fm} \ , \quad
\langle r^2 \rangle_{(1)2}^{1/2} = 0.896\,\mathrm{fm} \ .
\label{eq:rmstworadii} \eeq
The isovector radius is consistent with Refs.~\cite{BROWN86,FST97}.
At this time, we have no way to estimate errors for any of these
derived radii.

Equation (\ref{eq:rMnsquared}) also shows the relationships between
the Dirac and Pauli rms radii and the ${\cal O}(Q^2)$ parameters in
the EM coupling expansion (i.e., $\beta^{(0)}, \beta^{(1)}, f_v,
f_\rho$).  Simple inversion of these results allows us to present
parameters determined from the new single-nucleon data with the
nuclear couplings of set Q2:
\beq \beta^{(0)} = 0.0670 \ , \quad \beta^{(1)} = -0.2402 \ , \quad
f_v = -0.3279 \ , \quad f_\rho = 4.4198 \ .
\label{eq:newlowestorder}
\eeq

\section{Inclusion of Higher-Order Couplings}
\label{sec:newfit}

So far, we have adjusted the parameters that enter at
$\mathcal{O}(\tau )$ to achieve the modern values for the four rms
radii. Now we want to go to higher order in $\tau$ and see what
adjustments are necessary to reproduce the Sachs form factors up to
$Q\approx 600\,\mathrm{MeV}$ or $\tau \approx 0.1$.  We can relate
these adjustments to higher-order terms in the derivative expansion
of the EM lagrangian.

For the purposes of this work, we take the nucleon part of the EM
lagrangian as
\beqa \mathcal{L}_{\mathrm{EM}} &=& -e A_\mu \Nbar \gammamu
\frac{1}{2}(1 + \tau_3 )N -\frac{e}{4M} F_{\mu\nu} \Nbar \lambda
\sigmamunu N - \frac{e}{2M^2} (\partial^\nu F_{\mu\nu}) \Nbar \beta
\gammamu N \nonumber\\[4pt]
&\null& \quad {} -\frac{e}{4M^3} (\partial_\nu \partial^\eta
F_{\mu\eta}) \Nbar \lambda' \sigmamunu N - \frac{e}{M^4} (\partial^2
\partial^\nu F_{\mu\nu})\Nbar \beta' \gammamu N \nonumber\\[4pt]
&\null& \quad {} - \frac{e}{2M^5} (\partial^2 \partial_\nu
\partial^\eta F_{\mu\eta} ) \Nbar \lambda'' \sigmamunu N + \cdots \ ,
\label{eq:LEM} \eeqa
where $A^\mu$ and $F^{\mu\nu}$ are the electromagnetic fields. All
of the constants $\lambda, \beta, \lambda', \beta', \lambda''$ have
the isospin structure
\beqa \lambda &=& \lambda_p {1 \over 2} (1 + \tau_3 ) + \lambda_n {1
\over 2} (1 - \tau_3 ) = {1 \over 2} (\lambda_p + \lambda_n ) + {1
\over 2} \tau_3 (\lambda_p - \lambda_n ) \nonumber\\[4pt]
&\equiv& \lambda^{(0)} + \lambda^{(1)} \tau_3 \ , \quad
\mathrm{etc.}
\label{eq:isodef}
\eeqa
As shown in Ref.~\cite{EMQHD07}, the isovector parts of these
constants should be modified to include pion interactions to
maintain the residual chiral symmetry in the lagrangian.  When
applied to the nucleon form factors, however, these pions appear
only in loops, which we are not considering, so we have omitted
these terms.

We have already included the $\lambda$ and $\beta$ parameters in the
preceding section.  It is easy to see that the $\lambda'$ constants
enter the magnetic rms radii at the same order in $Q^2$ as the
vector meson couplings $f_v$ and $f_\rho$. Thus the $\lambda'$
parameters are redundant in our approach and will not be considered
in the sequel.  Thus the four new adjustable constants at our
disposal are contained in $\beta'$ and $\lambda''$.

It is straightforward to work out the Feynman rules for the new
vertices and to construct the tree-level contributions to the form
factors.  One finds that these constants enter the Dirac and Pauli
form factors at $\mathcal{O}(Q^4)$.  Thus they will enter the Sachs
form factors at $\mathcal{O}(\tau^2 )$ and $\mathcal{O}(\tau^3 )$.

Our strategy is the following: We begin with the magnetic form
factors, where both new terms enter at $\mathcal{O}(\tau^2 )$. We
numerically adjust the coefficient of this term until we get a good
fit to the form factor up to $\tau = 0.1$.  Here we define a good
fit as one that minimizes the maximum deviation between the fit and
the data (i.e., Kelly's results \cite{KELLY04}) throughout the whole
interval $0 \le \tau \le 0.1$. The result provides two constraints
between the $\beta'$ and $\lambda''$.  We then turn to the electric
form factors, where the new terms enter at both $\mathcal{O}(\tau^2
)$ and $\mathcal{O}(\tau^3 )$.  The coefficients of both of these
terms are then adjusted, consistent with the constraint, until a
good fit is obtained for $0 \le \tau \le 0.1$.  This procedure
provides us with four numbers that can be used to determine the
$\beta'$ and $\lambda''$.

\begin{figure}
\begin{center}
\includegraphics*[width=5.0in,angle=270]{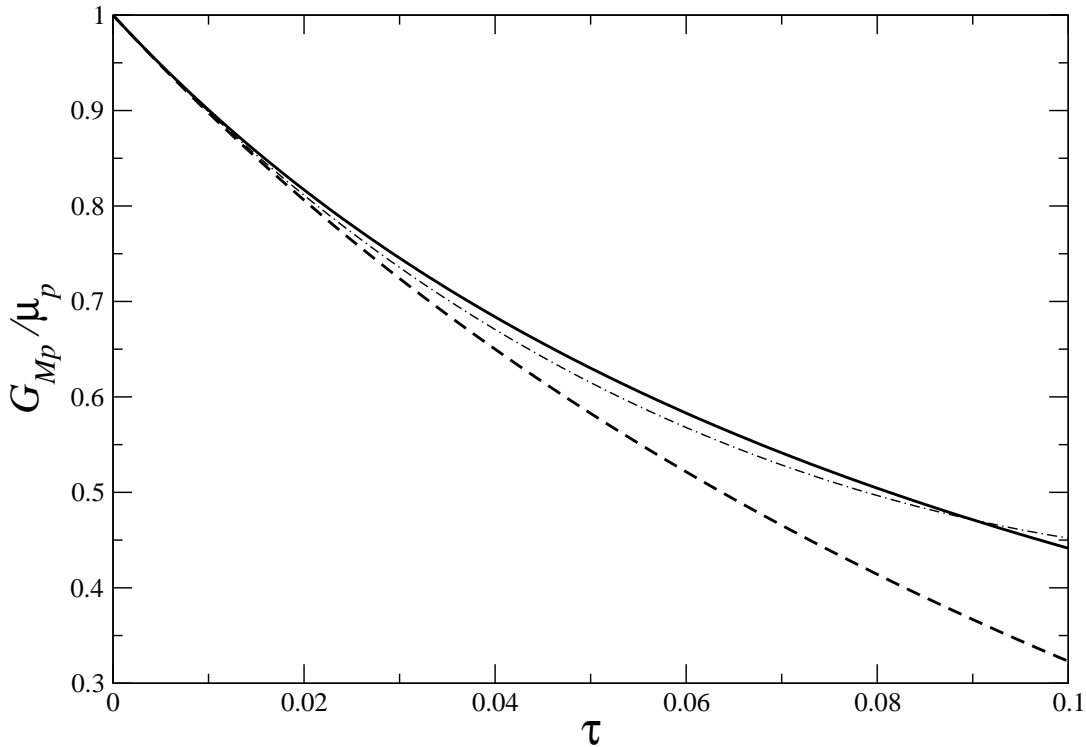}
\end{center}
 \caption{Proton magnetic form factor as a function of $\tau$. The
 curves represent the data (solid), the lowest-order fit (dashed),
 and the new fit (dot-dashed).}
 \label{fig:GMp}
\end{figure}

\begin{figure}
\begin{center}
\includegraphics*[width=5.0in,angle=270]{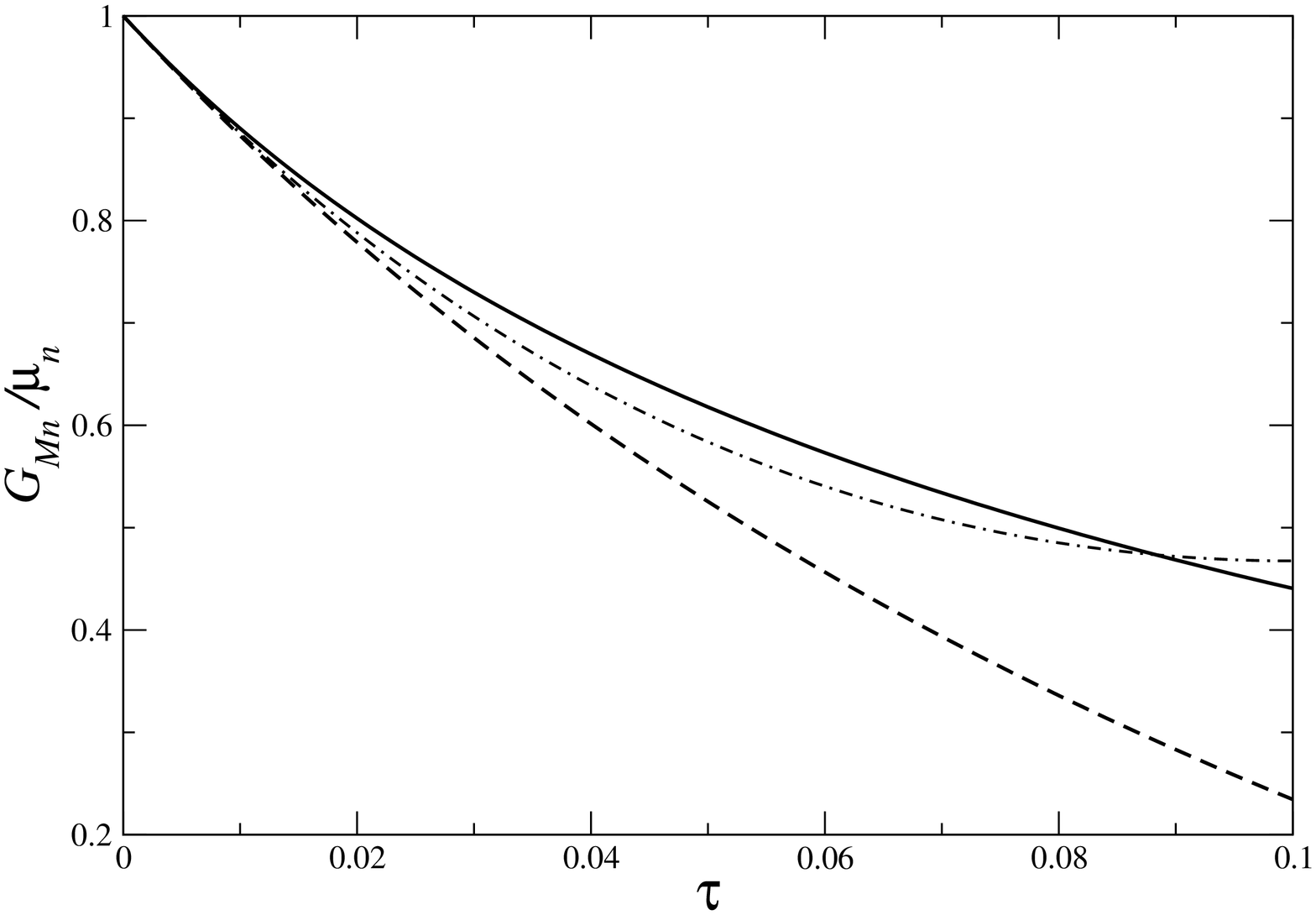}
\end{center}
 \caption{Neutron magnetic form factor as a function of $\tau$. The
 curves are identified as in Fig.~\protect\ref{fig:GMp}.}
 \label{fig:GMn}
\end{figure}

Denoting the new contributions to the Sachs form factors as $\delta
G$, we find
\beqa \delta \GMp (\tau) &=& 16 (\lambda''^{(0)} + \lambda''^{(1)} -
\beta'^{(0)} - \beta'^{(1)}) \tau^2 \ , \\[4pt]
\delta \GMn (\tau) &=& 16 (\lambda''^{(0)} - \lambda''^{(1)} -
\beta'^{(0)} + \beta'^{(1)}) \tau^2 \ . \label{eq:deltaGM} \eeqa
Taking the best fit to the data leads to Figs.~\ref{fig:GMp} and
\ref{fig:GMn}, and the constraints
\beq \lambda_p'' - \beta_p' = 2.255 \ , \quad \lambda_n'' - \beta_n'
= -2.786 \ .
\eeq
The fits are accurate to a few percent.

\begin{figure}
\begin{center}
\includegraphics*[width=5.0in,angle=270]{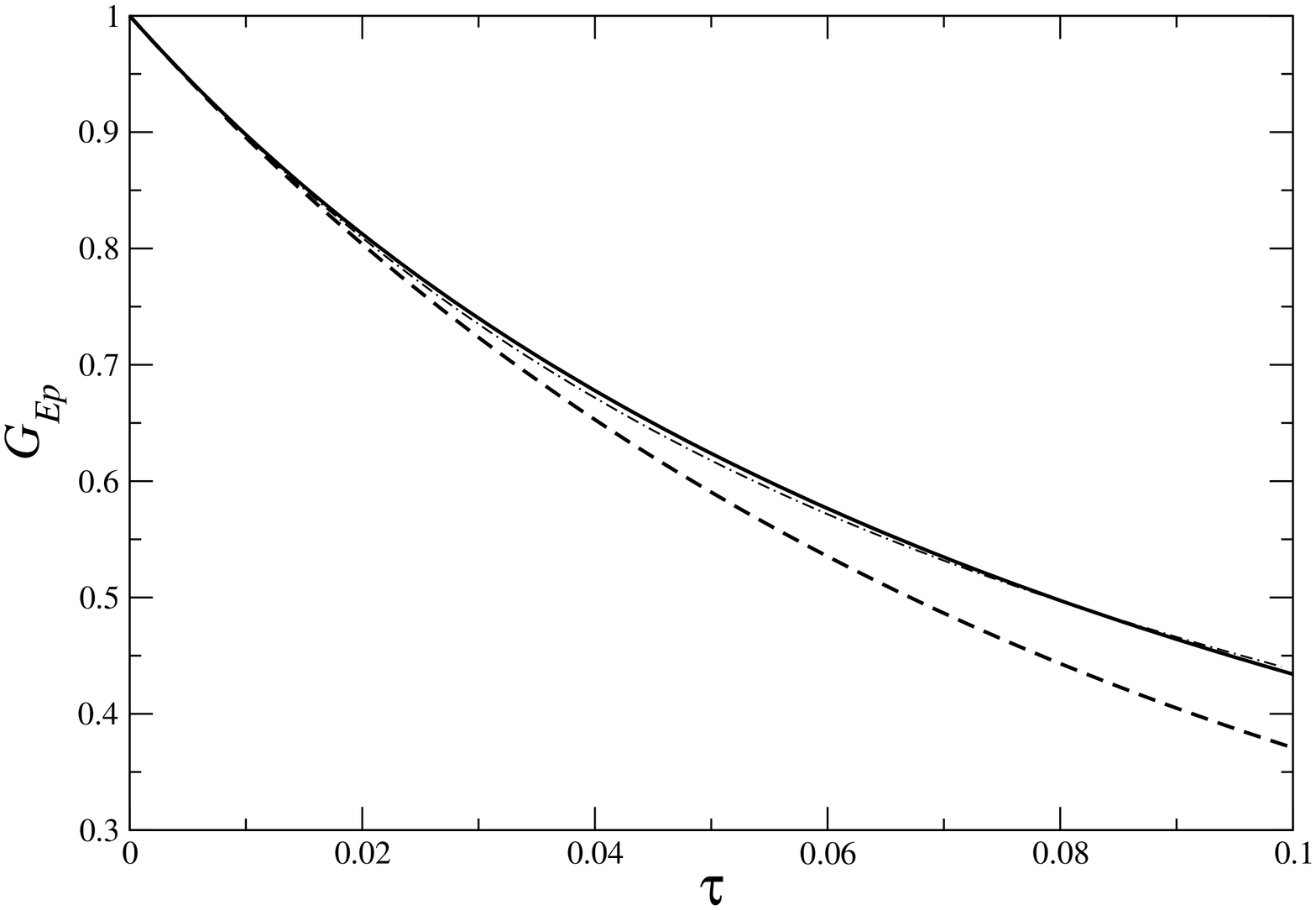}
\end{center}
 \caption{Proton electric form factor as a function of $\tau$. The
 curves are identified as in Fig.~\protect\ref{fig:GMp}.}
 \label{fig:GEp}
\end{figure}

\begin{figure}
\begin{center}
\includegraphics*[width=5.0in,angle=270]{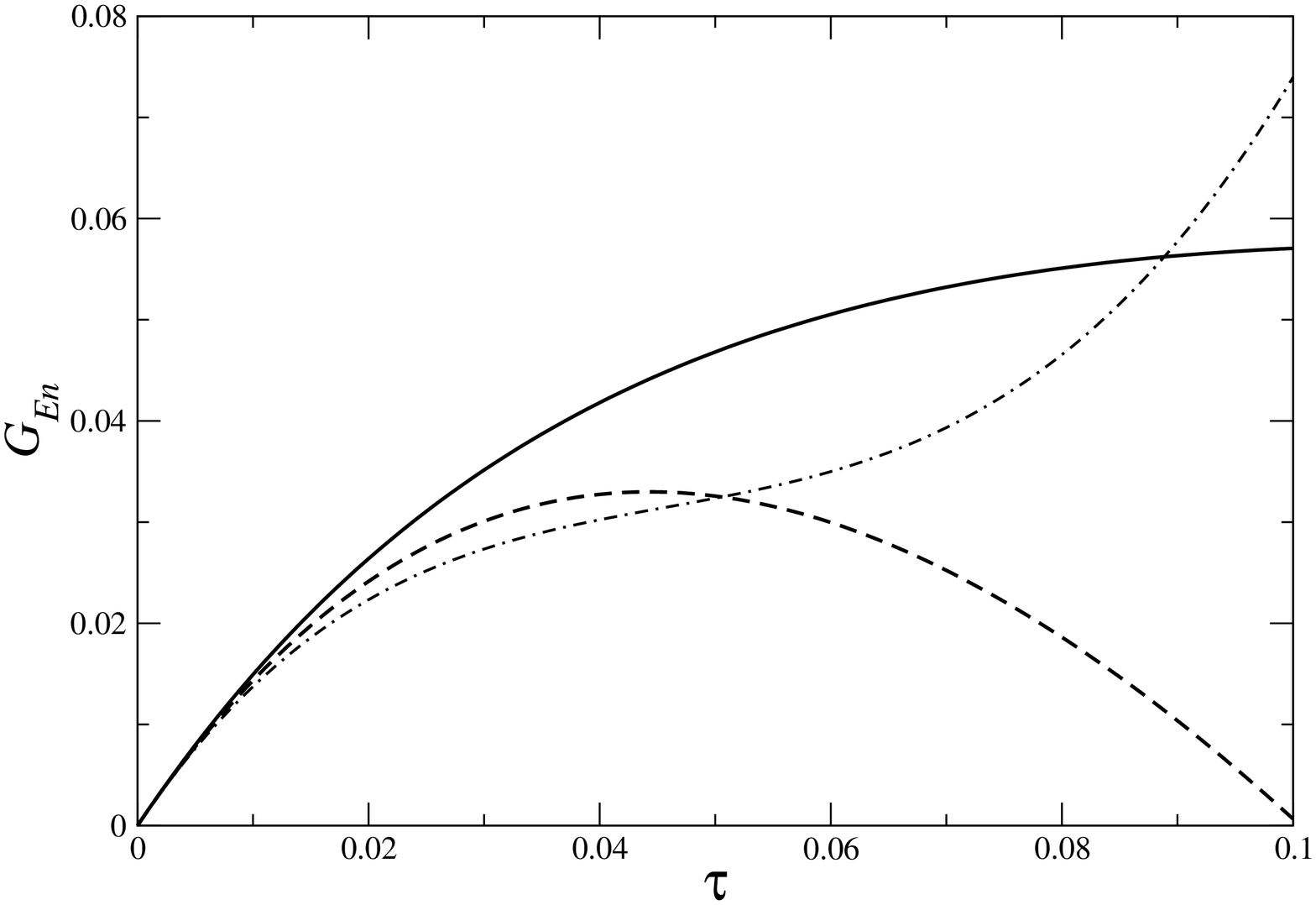}
\end{center}
 \caption{Neutron electric form factor as a function of $\tau$. The
 curves are identified as in Fig.~\protect\ref{fig:GMp}.}
 \label{fig:GEn}
\end{figure}

For the electric form factors, the new terms are
\beqa \delta \GEp (\tau) &=& -16(\beta'^{(0)} + \beta'^{(1)}) \tau^2
- 64 (\lambda''^{(0)} + \lambda''^{(1)}) \tau^3 \ ,\\[4pt]
\delta \GEn (\tau) &=& -16(\beta'^{(0)} - \beta'^{(1)}) \tau^2 - 64
(\lambda''^{(0)} - \lambda''^{(1)}) \tau^3 \  \label{eq:deltaGE}
\eeqa
Taking the best fit subject to the constraints above produces
Figs.~\ref{fig:GEp} and \ref{fig:GEn}, and the values
\beq \beta_p' = -0.9438 \ , \quad \beta_n' = 0.4688 \ . \eeq
Combining these results with the constraints determines all four new
parameters.

\begin{figure}
\begin{center}
\includegraphics*[width=5.0in,angle=270]{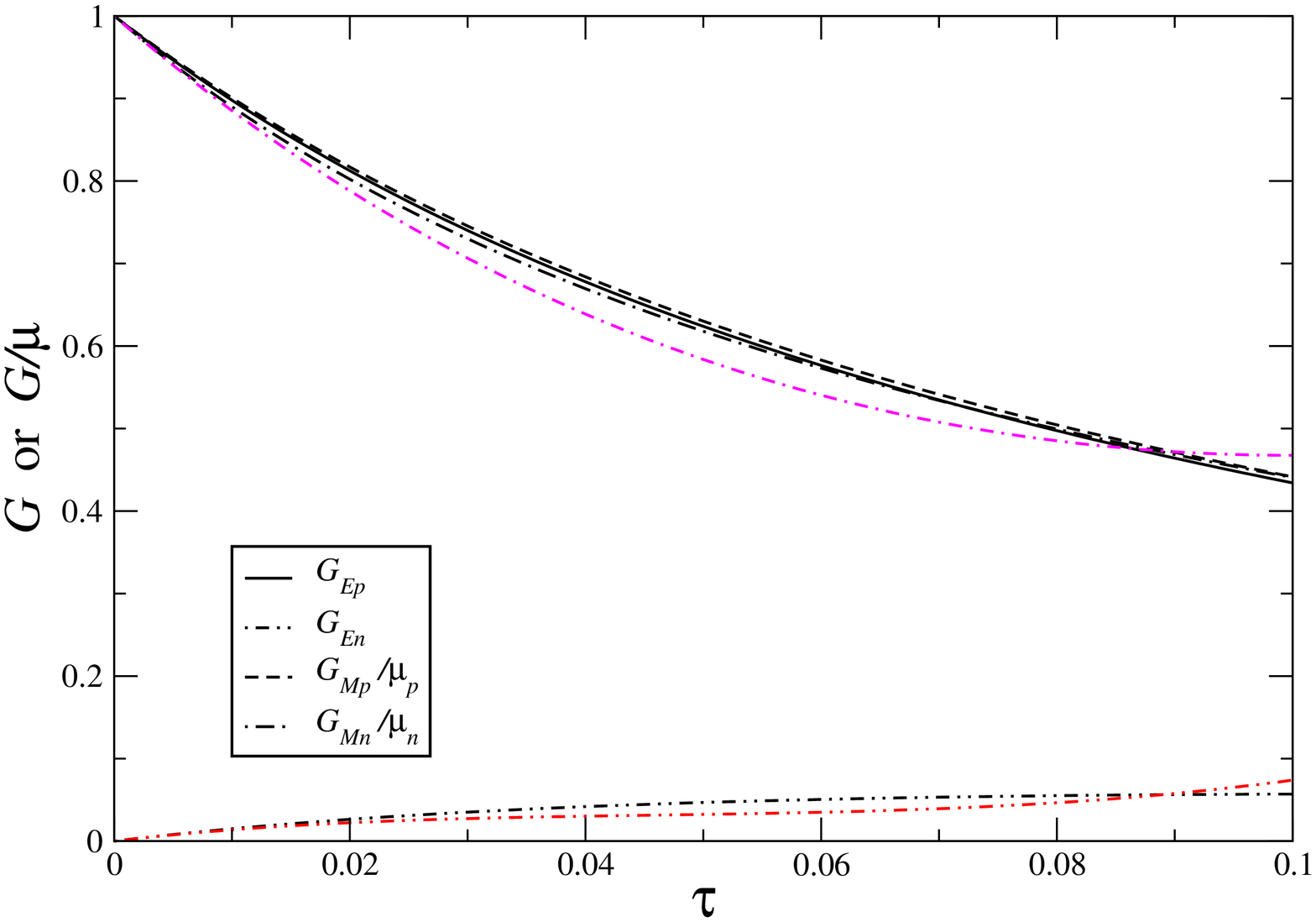}
\end{center}
 \caption{All four empirical form factors \protect\cite{KELLY04} and
 fits to $\GEn$ and $\GMn$.}
 \label{fig:gfits}
\end{figure}

Although the relative error in $\GEn$ is as large as 30\%, $\GEn$ is
small in the momentum-transfer range of interest.  A more relevant
comparison is given in Fig.~\ref{fig:gfits}, where all four of the
empirical form factors are shown along with fits to $\GEn$ and
$\GMn$.

\section{Summary}
\label{sec:summary}

In this paper we studied a field-theoretic parametrization of
single-nucleon EM form factors at low energy.  This parametrization
is part of a Lorentz-covariant, chiral invariant, hadronic effective
field theory that was proposed to study the nuclear many-body
problem.  We thus have a single lagrangian that describes the
nuclear structure, nuclear currents, and interaction vertices at low
energies. This is a natural framework for discussing the roles of
one-body and two-body currents in nuclear electromagnetic
interactions.

The parametrization of the form factors is based on a combination of
vector meson dominance and a derivative expansion for nucleon
interactions with the EM field.  At leading order in derivatives,
the form factors accurately reproduce the single-nucleon electron
scattering data only up to roughly 250 MeV momentum transfer.  To
study two-body exchange currents, however, one must reproduce the
form factors accurately up to at least 600 MeV momentum transfer.
This paper is proof of principle that by including the
next-to-leading order (nonredundant) derivatives, one can adequately
describe the form factors up to 600 MeV with our form of
parametrization.

We also revisited the leading-order parametrization and used a
modern data set to evaluate the expansion coefficients.  We found
that it is now possible to determine all four rms radii for the
Dirac and Pauli, isoscalar and isovector form factors, and thus
determine all four leading-order coefficients from the
single-nucleon data, unlike in Ref.~\cite{FST97}.

It is interesting that a simple power-series expansion in powers of
the momentum transfer squared would require many, many terms to
reproduce the form factors up to the desired 600 MeV.  (This is easy
to see using ten minutes of simple calculations on
\textsc{mathematica} with Kelly's \cite{KELLY04} fits to the data.)
Thus the vector meson dominance contributions are critical in
allowing us to parametrize the desired data using only the
next-to-leading order derivative terms.

While there is much work still to be done before two-body currents
can be studied numerically within this effective field theory
framework, we have shown that the one-body current can be adequately
parametrized to make a study of two-body currents possible.

\acknowledgments

This work was supported in part by the Department of Energy under
Contract No.\ DE--FG02--87ER40365.

\end{document}